\documentclass[english,aps,prx,twocolumn,superscriptaddress,longbibliography]{revtex4-2}
\usepackage{amssymb}
\usepackage{amsmath}
\usepackage{graphicx}
\usepackage{natbib}
\usepackage{epstopdf}
\usepackage{color}
\usepackage{braket}
\usepackage{physics}
\usepackage{mathrsfs}
\usepackage{upgreek}
\usepackage{todonotes}
\usepackage[colorlinks=true, allcolors=blue]{hyperref}

\begin{document}
\title{Dispersive readout of a silicon quantum device using an atomic force microscope-based rf gate sensor}

\author{Artem O. Denisov}
\email{adenisov@princeton.edu}
\affiliation{Department of Physics, Princeton University, Princeton, New Jersey 08544, USA}
\author{Gordian Fuchs}
\affiliation{Department of Physics, Princeton University, Princeton, New Jersey 08544, USA}
\author{Seong W. Oh}
\altaffiliation{Present Address: Department of Electrical and Systems Engineering, University of Pennsylvania, Philadelphia, Pennsylvania~19104, USA}
\affiliation{Department of Physics, Princeton University, Princeton, New Jersey 08544, USA}
\author{Jason R. Petta}
\email{petta@physics.ucla.edu}
\affiliation{Department of Physics, Princeton University, Princeton, New Jersey 08544, USA}
\affiliation{Department of Physics and Astronomy, University of California, Los Angeles, California 90095, USA}
\affiliation{Center for Quantum Science and Engineering, University of California, Los Angeles, California 90095, USA}

\begin{abstract}
We demonstrate dispersive charge sensing of Si/SiGe single and double quantum dots (DQD) by coupling sub-micron floating gates to a radio frequency reflectometry (rf-reflectometry) circuit using the tip of an atomic force microscope (AFM). Charge stability diagrams are obtained in the phase response of the reflected rf signal. We demonstrate single-electron dot-to-lead and dot-to-dot charge transitions with a signal-to-noise ratio (SNR) of $2$ and integration time of $\tau~=~2.7~\mathrm{ms}$ and $\tau~=~6.4~\mathrm{ms}$, respectively. The charge sensing SNR compares favorably with results obtained on conventional devices. Moreover, the small size of the floating gates largely eliminates the coupling to parasitic charge traps that can complicate the interpretation of the dispersive charge sensing data.
\end{abstract}

\maketitle
\section{Introduction}
Many quantum phenomena are only accessible through time-resolved measurements with short averaging times. A ubiquitous example is a projective measurement of a two-level system, where the measurement must be performed before the excited state can relax to the ground state~\cite{Wallraff2004, Elzerman2004, doi:10.1126/science.1116955}. The rapid development of quantum information technologies in solid state systems has in part been enabled by major breakthroughs in readout, such as rf-reflectometry~\cite{2202.10516, doi:10.1063/1.2794995} and dispersive measurements in the circuit quantum electrodynamics (cQED) device architecture~\cite{RevModPhys.93.025005}. Future implementations of quantum algorithms~\cite{nielsen_chuang} and quantum error correcting protocols~\cite{quant-ph/9712048} will be heavily dependent on accurate quantum state measurements. 

\begin{figure}[tbh!]
	\centering
	\includegraphics[width=1\columnwidth]{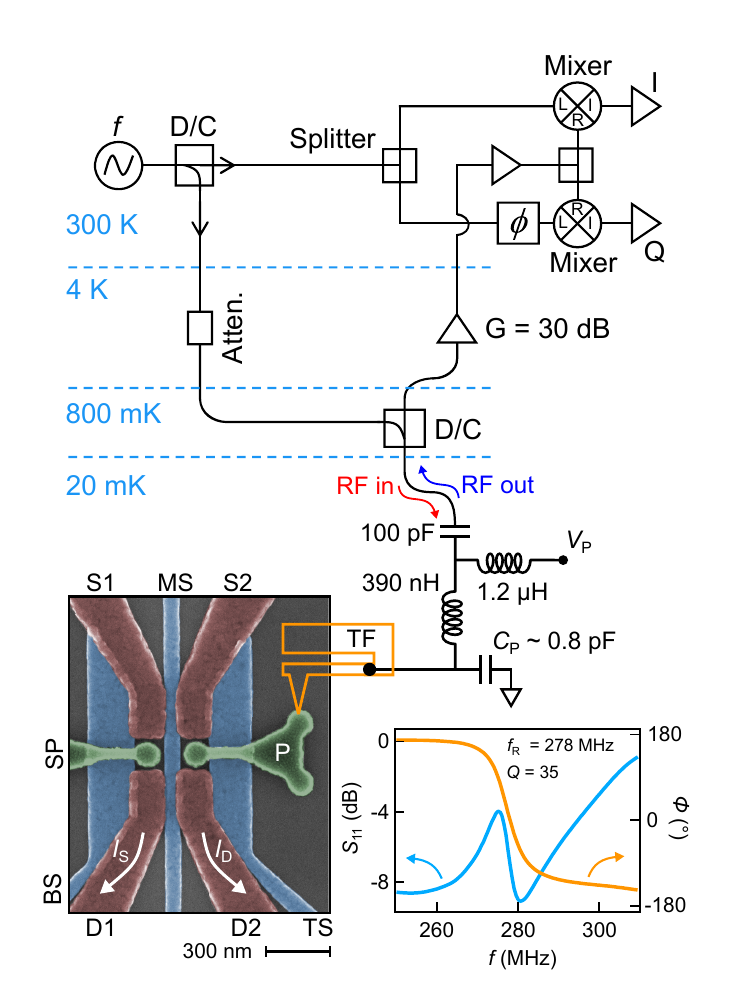}
	\caption{Measurement circuit diagram and false-color scanning electron microscope (SEM) image of the device. The overlapping gate design incorporates Al screening (blue), Al accumulation (red), and Pd plunger gates (green). The floating Y-shaped plunger gate (P) is coupled to the $LC$ circuit using an AFM tip located on the prong of a quartz tuning fork (TF). $I_{\mathrm{D}}$ refers to the current through the QD that is coupled to the AFM tip. $I_{\mathrm{S}}$ is the current through the conventional dc charge sensing QD. Inset: Amplitude $S_{11}$ and phase $\phi$ response of the resonant circuit measured as a function of drive frequency $f$ at a base temperature of $20~\mathrm{mK}$.}
	\label{fig_1}
\end{figure}

Spin qubits in semiconductor quantum dots (QDs) are one of the most promising platforms for the practical implementation of large-scale quantum information processors~\cite{JRP_review}. Long coherence times~\cite{Tyryshkin2012}, high two-qubit gate fidelities ~\cite{doi:10.1126/sciadv.abn5130, Xue2022, Noiri2022}, and compatibility with industrial fabrication techniques~\cite{Maurand2016, Zwerver2022, Xue2021} have all been demonstrated. Charge sensing is a crucial technique for measuring spin qubits, as it provides access to the spin degree-of-freedom through spin-to-charge conversion~\cite{Elzerman2004, PhysRevLett.103.160503, doi:10.1126/science.1116955}. Typically, sensitive charge readout is achieved by locating a quantum point contact (QPC) or sensing QD in the vicinity of the qubit~\cite{RevModPhys.79.1217, JRP_review}. rf readout of charge detectors has been implemented by coupling them to impedance-matched resonant circuits~\cite{doi:10.1126/science.280.5367.1238, doi:10.1063/1.2794995}. Alternatively, it is feasible to leverage rf-reflectometry for readout without the additional overhead of separate QPCs and sensing QDs~\cite{2202.10516}. This approach, termed dispersive gate sensing, couples the gates that define the electrostatic potential of the QD directly to a $LC$ resonant circuit. Dispersive gate sensing was first implemented with GaAs QDs~\cite{PhysRevLett.110.046805} and later extended to silicon-on-insulator (SOI)~\cite{Crippa2019, doi:10.1021/acs.nanolett.5b04356, doi:10.1021/acs.nanolett.5b01306} and accumulation-mode gate-defined devices~\cite{Rossi, West2019, Zheng2019}. Single-shot spin state readout has recently been achieved using dispersive gate sensing~\cite{West2019, Zheng2019, PhysRevX.8.041032, Urdampilleta2019}.

\begin{figure}[t!]
	\centering
	\includegraphics[width=\columnwidth]{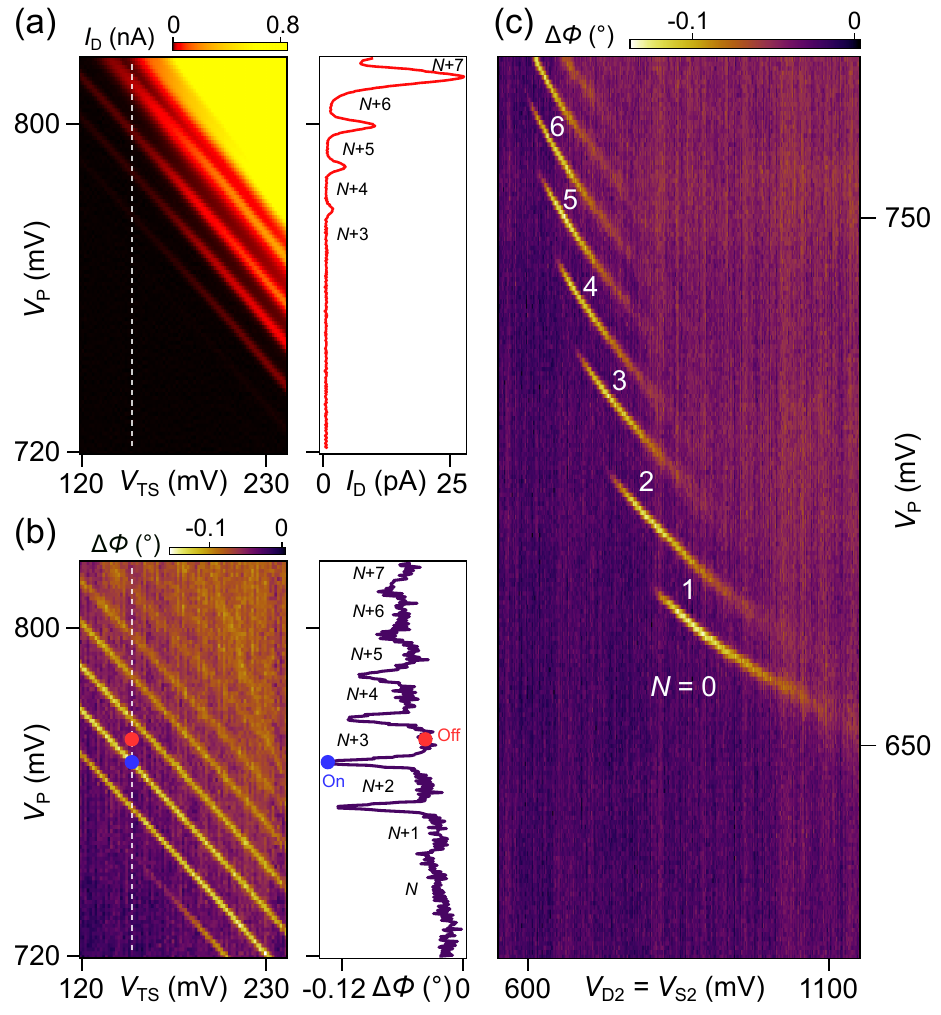}
	\caption{(a) Simultaneously measured current $I_{\mathrm{D}}$ through the QD and (b) phase response $\Delta \phi$ plotted as a function of $V_{\mathrm{P}}$ and $V_{\mathrm{TS}}$. Linecuts across white dashed lines are shown to the right. Blue and red points indicate on-peak and off-peak states used for signal-to-noise ratio analysis. (c) Phase response in the single electron regime. The dispersive shift is shown as a function of plunger $V_{\mathrm{P}}$ and combined $V_{\mathrm{D2}}=V_{\mathrm{S2}}$ accumulation gates voltages. The number of electrons in the dot is denoted by $N$. }
	\label{fig_2}
\end{figure}

In this Letter, we couple a rf-reflectometry circuit to the tip of a milli-Kelvin AFM~\cite{Seong_AIP, doi:10.1021/acs.nanolett.2c01098}. We demonstrate dispersive charge readout of Si/SiGe single QD and DQD devices by electrically connecting the AFM tip to sub-micron floating plunger gates fabricated on the surface of a Si/SiGe heterostructure~\cite{2302.07949}. For the DQD device, we compare standard dc charge sensing measurements with the resonator phase response to confirm single electron occupation. A charge sensing SNR of unity is achieved with a few $\mathrm{ms}$ integration time, comparable to SNRs obtained on conventional devices~\cite{West2019}.

\section{Experimental setup}
Measurements are performed on QD devices fabricated on an undoped Si/SiGe heterostructure with a $5~\mathrm{nm}$ thick Si quantum well (QW). We utilize a gate stack consisting of three overlapping layers to form a QD in the plane of the QW~\cite{PhysRevApplied.6.054013}. For the device shown in Fig.~\ref{fig_1}, two Al layers form the screening (MS, BS, TS) and accumulation gates (S1, S2, D1, and D2), while the Pd Y-shaped island on top acts as a floating plunger gate (P)~\cite{2302.07949}. We first measure a single QD defined under the floating P-gate in Fig.~\ref{fig_1}. The device was located using the AFM in topographic mode and then the tip was brought into galvanic contact with the floating gate. To control the occupancy of the QD, a dc voltage $V_{\mathrm{P}}$ is applied to the tip, while the tunnel coupling is controlled by accumulation gates S2 and D2. Screening gates MS and TS define the channel where the source/drain reservoirs and QD are formed.

\begin{figure*}[tbh!]
	\centering
	\includegraphics[width=1.85\columnwidth]{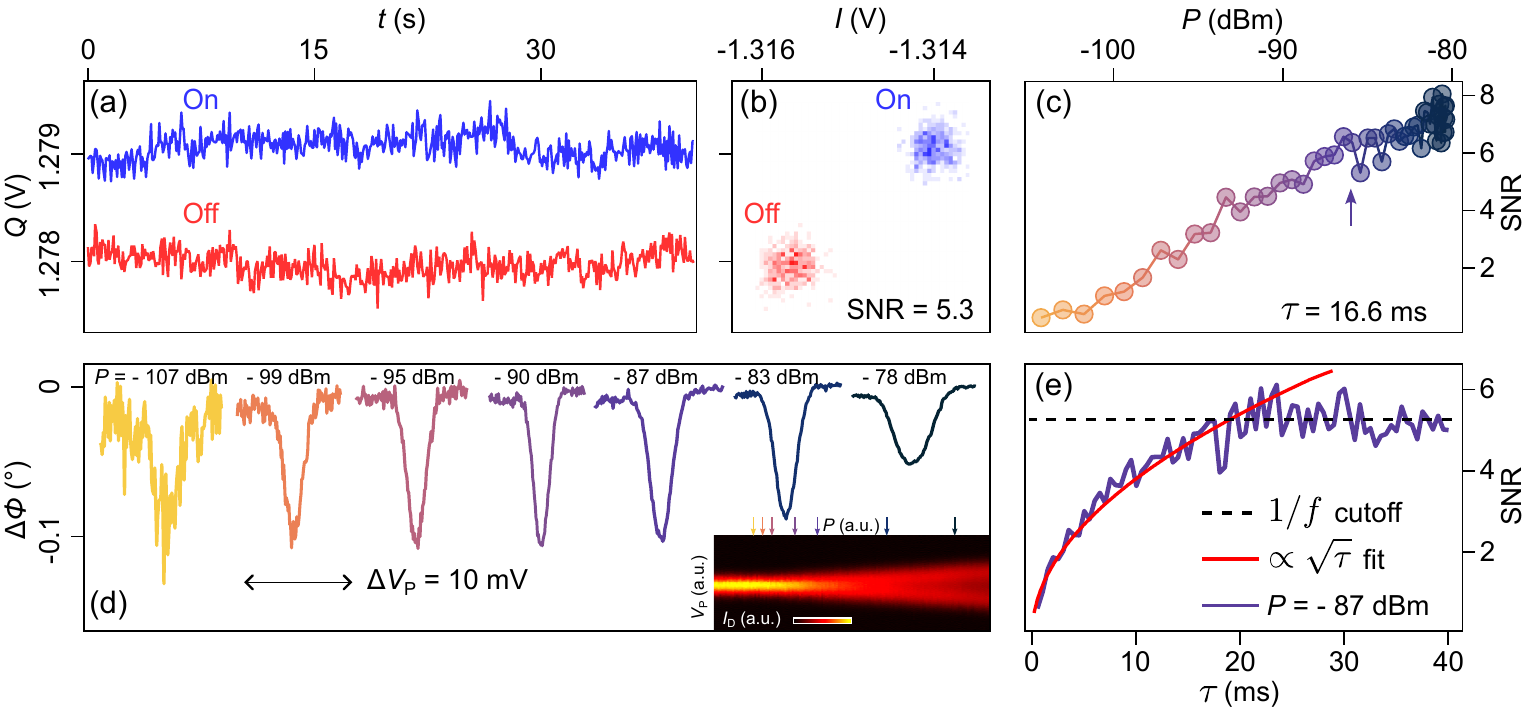}
	\caption{Signal-to-noise ratio (SNR) analysis. (a) Time trace of the demodulated $Q$ quadrature of the reflected signal. On-state (blue) is sampled at the dip of the phase response, right in the middle of the charge transition as highlighted by the blue dot in Fig.~\ref{fig_2}(b). Off-state (red) is sampled in Coulomb blockade as highlighted by the red dot in Fig.~\ref{fig_2}(b). (b) Combined histogram of the $I$ and $Q$ components obtained with a sampling rate of $240~\mathrm{kHz}$ and $\tau\sim16.7~\mathrm{ms}$. (c) SNR as a function of $P$ applied at the resonator. The arrow indicates the optimal power $P=-87~\mathrm{dBm}$. (d) Phase response around the charge transition as a function of $V_{\rm P}$ for increasing $P$. Curves are color-matched to the SNR data in (c). Inset: Evolution of the corresponding Coulomb blockade peak as measured in the transport current $I_{\rm D}$ as a function of $P$. Arrows are color-matched to the phase response curves and highlight the corresponding drive powers. (e) SNR as a function of $\tau$ with $P=-87~\mathrm{dBm}$. The expected square-root dependence is depicted by the red curve and the $1/f$ noise saturation value is marked by the dashed line. }
	\label{fig_3}
\end{figure*}

Charge detection is achieved by coupling the rf-reflectometry circuit shown in Fig.~\ref{fig_1} directly to the AFM tip. A directional coupler (D/C) splits the rf drive signal into two paths, one which propagates down to the AFM tip and another used as the local oscillator (LO) for homodyne detection. At the mK temperature stage of the dilution refrigerator, a portion of the drive signal is reflected from a tank circuit consisting of a surface mount inductor $L~=~390~\mathrm{nH}$ and a parasitic capacitance $C_{\mathrm{p}}\sim0.8~\mathrm{pF}$. As the phase response $\Delta \Phi~\approx~-2 Q \Delta C/C_{\mathrm{p}}$ is directly related to the change in the total system capacitance $\Delta C$, it is crucial to minimize $C_{\mathrm{p}}$~\cite{2202.10516}. In order to do so, all circuit components are mounted directly to the AFM sensor holder, in close proximity to the TF~\cite{Seong_AIP}. The reflected signal is cryogenically amplified at the $4~\mathrm{K}$ plate before being further amplified at room temperature~\cite{doi:10.1063/1.3103939}. A power splitter, phase shifter, and two mixers enable the simultaneous acquisition of the rf amplitude and phase response~\cite{doi:10.1063/1.4729469}. A surface mounted bias tee allows rf and dc voltages to be coupled to the AFM tip. The inset of Fig.~\ref{fig_1} shows the amplitude $S_{\mathrm{11}}$ and phase $\phi$ response of the rf-reflectometry circuit measured as a function of drive frequency $f$, yielding a resonance frequency $f_{\mathrm{r}}~=~278.5~\mathrm{MHz}$ and quality factor $Q~=~35$. The readout bandwidth is $f_{r}/Q~\approx~8~\mathrm{MHz}$.

\section{Single dot dispersive charge sensing}
We first measure a single QD defined under the P floating gate in Fig.~\ref{fig_1}. Coulomb blockade peaks are visible in the dot current, $I_{\rm D}$, which is measured as a function of $V_{\mathrm{P}}$ and $V_{\mathrm{TS}}$ in Fig.~\ref{fig_2}(a). The resonator phase response, $\Delta \phi$, exhibits features at charge degeneracy lines [Fig.~\ref{fig_2}(b)]. Near the charge degeneracy lines, the electron tunneling rate between the QD and source/drain reservoirs is faster than the rf modulation frequency $f_{\mathrm{R}}$, thus effectively contributing to the total capacitance (tunneling capacitance) of the tank circuit $\Delta C = (\alpha e)^2/4k_{\mathrm{B}}T$, where $\alpha e$~is the lever arm of the plunger gate ($\sim0.1~\mathrm{eV/V}$), $k_{\mathrm{B}}$ is Boltzmann's constant and $T$ is the reservoir electron temperature~\cite{2202.10516}. Electron tunneling is prohibited in Coulomb blockade and the phase remains nearly constant elsewhere. Line cuts along the plunger voltage axis, $V_{\mathrm{P}}$, clearly show that dispersive sensing provides information about the charge state of the QD even when the dc current through the device is too small to be measured.

We demonstrate dispersive charge sensing down to the last electron in Fig.~\ref{fig_2}(c). Here, the phase shift is shown as a function of plunger $V_{\mathrm{P}}$ and combined barrier gate $V_{\mathrm{D2}}=V_{\mathrm{S2}}$ voltages. The latter are used to control the tunneling rates to the leads. The lowest feature in the data corresponds to the  $N$ = 0 $\leftrightarrow$ 1 charge transition in the QD. As the voltage of the barrier gates decreases, the transition lines fade and eventually disappear when the tunneling rate becomes much slower than the rf modulation frequency and no effective electron exchange with the reservoir is possible. At high barrier voltages, the tunneling rate becomes large and the dispersive peaks smear due to lifetime broadening. It should be noted, that owing to our compact floating gate design, we do not observe parasitic background fluctuations in the dispersive signal~\cite{Rossi, West2019}.

\section{Charge sensing SNR}
We optimize the performance of our system as a function of rf drive power $P$ and integration time $\tau$. To evaluate the signal-to-noise ratio (SNR), we sample the $I$ and $Q$ quadrature signals at the bottom of the phase dip (on-state) and deep in Coulomb blockade (off-state),  as illustrated by the red and blue dots in Fig.~\ref{fig_2}(b). A typical time trace of the $Q$ quadrature acquired with a sampling rate of $240~{\mathrm{kHz}}$ and $\tau~=~16.7~\mathrm{ms}$ is shown in Fig.~\ref{fig_3}(a). We estimate the SNR as the separation of the on- and off-states divided by the standard deviation~\cite{West2019}:
\begin{equation}
\mathrm{SNR}=\frac{\langle I_{\mathrm{on}}-I_{\mathrm{off}} \rangle^2+\langle Q_{\mathrm{on}}-Q_{\mathrm{off}} \rangle^2}{\mathrm{std}[( I_{\mathrm{on}}-I_{\mathrm{off}} )^2+(Q_{\mathrm{on}}-Q_{\mathrm{off}} )^2]}.	
\end{equation}
$IQ$-histograms acquired in the on- and off-states are shown in Fig.~\ref{fig_3}(b) and yield a SNR~$=5.3$. The SNR increases as we apply more power [Fig.~\ref{fig_3}(c)]. Figure~\ref{fig_3}(d) shows how the charge transition also broadens with increasing $P$. We achieve an optimal SNR with $P\sim-87~\mathrm{dBm}$. Finally, we measure the SNR as a function of $\tau$ at the optimal power [Fig.~\ref{fig_3}(e)]. For $\tau \approx 2.7~\mathrm{ms}$ we can detect the charge transition with a SNR~$=~2$. The expected square root behavior is observed for $\tau \lesssim 25~\mathrm{ms}$. At higher integration times, the SNR saturates around a value of~$\sim 5$ meaning that longer averaging does not improve the SNR due to low frequency charge noise.

\section{Double quantum dot dispersive charge sensing}

\begin{figure}[t!]
	\centering
	\includegraphics[width=\columnwidth]{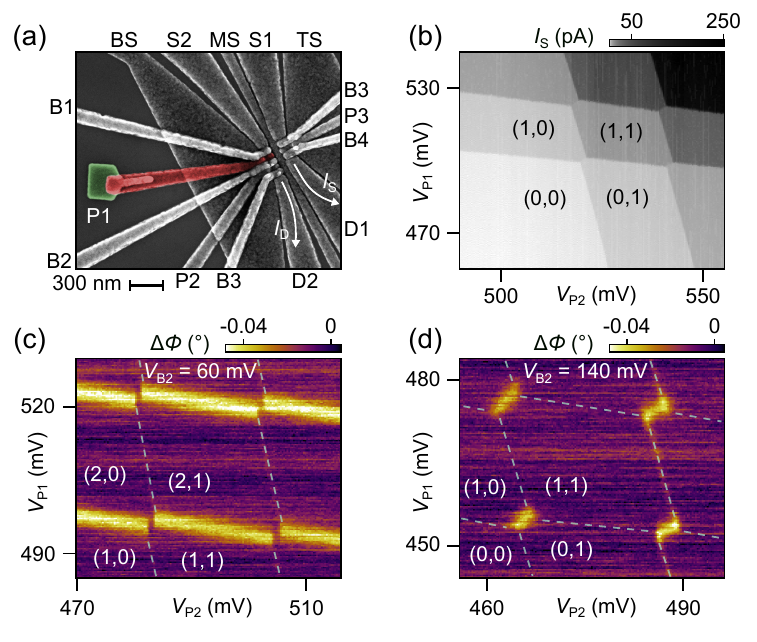}
	\caption{(a) False-color SEM image of the DQD device accompanied by a conventional dc charge sensing dot. The floating Al plunger gate P1 is highlighted in red. The Pd AFM tip contact is highlighted in green. (b) Charge stability diagram in the single electron regime acquired by measuring the current $I_{\mathrm{S}}$ through the charge sensing dot. (c) Measured dispersive response as a function of $V_{\rm P1}$ and $V_{\rm P2}$ in a regime where the dots are weakly coupled. The charge occupation was confirmed by conventional charge sensing measurements (not shown). (d) Dispersive response in a regime with strong interdot tunnel coupling.}
	\label{fig_4}
\end{figure}

Finally, we demonstrate AFM-based dispersive charge sensing of a DQD. Figure~\ref{fig_4}(a) shows a false-color SEM image of the device. In contrast to the single QD device from Fig.~\ref{fig_1}, this device is fabricated using three overlapping Al gate layers~\cite{PhysRevApplied.6.054013}. In order to create good electrical contact between the AFM tip and the floating gate P1, we pre-patterned a small $300\times200~\mathrm{nm}$ Pd pad highlighted in green in the SEM image. 

Conventional charge sensing using a separate sensing QD or QPC allows the unambiguous determination of the electronic occupation ($N_1$,$N_2$) of the DQD even when the tunneling rate to the leads is pinched off down to the kHz regime. In contrast, rf dispersive readout is only sensitive to tunneling rates that are faster than the excitation frequency. Figure~\ref{fig_4}(b) shows a charge stability diagram obtained by measuring the dc current $I_{\mathrm{S}}$ through the sensor dot as a function of $V_{\rm P1}$ and $V_{\rm P2}$. The plunger gate P1 is voltage-biased using the AFM tip.

We measure charge stability diagrams by simultaneously acquiring the dispersive and dc-charge sensor (not shown) response. We begin in a regime where the two QDs are weakly coupled ($V_{\mathrm{B2}}=60~\mathrm{mV}$). Figure~\ref{fig_4}(c) shows the dispersive shift as a function of $V_{\rm P1}$ and $V_{\rm P2}$. The charge stability diagram around the (2,0)-(1,1) transition is clearly visible in the phase response, which was confirmed independently by the charge sensing dot. Here, the QD formed under the P1 gate serves as a single electron box (SEB) rf charge detector that is coupled only to one reservoir~\cite{Ansaloni2020}. For larger middle barrier voltages $V_{\mathrm{B2}}=140~\mathrm{mV}$, the stability diagram evolves into a honeycomb pattern as plotted in Fig.~\ref{fig_4}(d). The dot-to-lead transitions here are not visible due to slow tunneling rates. However, at the interdot transitions, we detect a phase shift $\Delta \Phi\approx-0.04^{\circ}$ related to the additional interdot tunneling quantum capacitance $\Delta C_{\mathrm{Q}} = (\alpha e)^2/4t_{\mathrm{c}}$, where $t_{\mathrm{c}}$ is the interdot tunnel coupling~\cite{2202.10516}. For (0,1)-(1,0) and higher transitions we achieve a SNR$=2$ with $\tau\approx 6.4~\mathrm{ms}$, which is comparable to conventional accumulation-mode systems. West \textit{et al.} recently demonstrated gate-based single-shot spin readout with a SNR = 2 for $\tau$ = 12 ms on the interdot transition in silicon MOS device~\cite{West2019}.

\section{Conclusion}
In conclusion, we have demonstrated dispersive gate-based charge sensing using a milli-Kelvin AFM with an integrated rf-tank circuit. The charge sensing SNR and readout time compare favorably with conventional dispersive gate sensing approaches. Previous dispersive gate sensing measurements in gate-defined devices were complicated by the rf response of background charges induced under the sensor-gates~\cite{Rossi, West2019}. The background charges can significantly contribute to $C_{\mathrm{p}}$ even if they are far away from the QD~\cite{PhysRevApplied.11.064027}. By reducing the gate footprint down to a sub-micron island, we vastly minimize electron accumulation under the sensing gate and minimize its geometric contribution to the parasitic capacitance. Additionally, the floating gate design combined with AFM-based dispersive readout enables the characterization of multiple devices on a single chip without the need for a complex multiplexing setup~\cite{2202.10516, doi:10.1063/1.4868107}. The ability to form and read out the charge state of a QD using an AFM tip opens up an exciting avenue of creating movable QDs in Si/SiGe heterostructures. 
\\
\begin{acknowledgments}
Supported by Army Research Office grant W911NF-15-1-0149 and the Gordon and Betty Moore Foundation’s EPiQS Initiative through Grant No.~GBMF4535. The authors thank F.~Ansaloni and F.~Borjans for useful discussions, F.~Borjans and A.~Mills for technical contributions to device fabrication, and HRL Laboratories for providing the Si/SiGe heterostructures.
\end{acknowledgments}

\bibliographystyle{apsrev4-2}
\bibliography{bib_Denisov_APL_v3}

\begin{thebibliography}{37}%
\makeatletter
\providecommand \@ifxundefined [1]{%
 \@ifx{#1\undefined}
}%
\providecommand \@ifnum [1]{%
 \ifnum #1\expandafter \@firstoftwo
 \else \expandafter \@secondoftwo
 \fi
}%
\providecommand \@ifx [1]{%
 \ifx #1\expandafter \@firstoftwo
 \else \expandafter \@secondoftwo
 \fi
}%
\providecommand \natexlab [1]{#1}%
\providecommand \enquote  [1]{``#1''}%
\providecommand \bibnamefont  [1]{#1}%
\providecommand \bibfnamefont [1]{#1}%
\providecommand \citenamefont [1]{#1}%
\providecommand \href@noop [0]{\@secondoftwo}%
\providecommand \href [0]{\begingroup \@sanitize@url \@href}%
\providecommand \@href[1]{\@@startlink{#1}\@@href}%
\providecommand \@@href[1]{\endgroup#1\@@endlink}%
\providecommand \@sanitize@url [0]{\catcode `\\12\catcode `\$12\catcode
  `\&12\catcode `\#12\catcode `\^12\catcode `\_12\catcode `\%12\relax}%
\providecommand \@@startlink[1]{}%
\providecommand \@@endlink[0]{}%
\providecommand \url  [0]{\begingroup\@sanitize@url \@url }%
\providecommand \@url [1]{\endgroup\@href {#1}{\urlprefix }}%
\providecommand \urlprefix  [0]{URL }%
\providecommand \Eprint [0]{\href }%
\providecommand \doibase [0]{https://doi.org/}%
\providecommand \selectlanguage [0]{\@gobble}%
\providecommand \bibinfo  [0]{\@secondoftwo}%
\providecommand \bibfield  [0]{\@secondoftwo}%
\providecommand \translation [1]{[#1]}%
\providecommand \BibitemOpen [0]{}%
\providecommand \bibitemStop [0]{}%
\providecommand \bibitemNoStop [0]{.\EOS\space}%
\providecommand \EOS [0]{\spacefactor3000\relax}%
\providecommand \BibitemShut  [1]{\csname bibitem#1\endcsname}%
\let\auto@bib@innerbib\@empty
\bibitem [{\citenamefont {Wallraff}\ \emph {et~al.}(2004)\citenamefont
  {Wallraff}, \citenamefont {Schuster}, \citenamefont {Blais}, \citenamefont
  {Frunzio}, \citenamefont {Huang}, \citenamefont {Majer}, \citenamefont
  {Kumar}, \citenamefont {Girvin},\ and\ \citenamefont
  {Schoelkopf}}]{Wallraff2004}%
  \BibitemOpen
  \bibfield  {author} {\bibinfo {author} {\bibfnamefont {A.}~\bibnamefont
  {Wallraff}}, \bibinfo {author} {\bibfnamefont {D.~I.}\ \bibnamefont
  {Schuster}}, \bibinfo {author} {\bibfnamefont {A.}~\bibnamefont {Blais}},
  \bibinfo {author} {\bibfnamefont {L.}~\bibnamefont {Frunzio}}, \bibinfo
  {author} {\bibfnamefont {R.-.~S.}\ \bibnamefont {Huang}}, \bibinfo {author}
  {\bibfnamefont {J.}~\bibnamefont {Majer}}, \bibinfo {author} {\bibfnamefont
  {S.}~\bibnamefont {Kumar}}, \bibinfo {author} {\bibfnamefont {S.~M.}\
  \bibnamefont {Girvin}},\ and\ \bibinfo {author} {\bibfnamefont {R.~J.}\
  \bibnamefont {Schoelkopf}},\ }\href {https://doi.org/10.1038/nature02851}
  {\bibfield  {journal} {\bibinfo  {journal} {Nature}\ }\textbf {\bibinfo
  {volume} {431}},\ \bibinfo {pages} {162} (\bibinfo {year}
  {2004})}\BibitemShut {NoStop}%
\bibitem [{\citenamefont {Elzerman}\ \emph {et~al.}(2004)\citenamefont
  {Elzerman}, \citenamefont {Hanson}, \citenamefont {Willems~van Beveren},
  \citenamefont {Witkamp}, \citenamefont {Vandersypen},\ and\ \citenamefont
  {Kouwenhoven}}]{Elzerman2004}%
  \BibitemOpen
  \bibfield  {author} {\bibinfo {author} {\bibfnamefont {J.~M.}\ \bibnamefont
  {Elzerman}}, \bibinfo {author} {\bibfnamefont {R.}~\bibnamefont {Hanson}},
  \bibinfo {author} {\bibfnamefont {L.~H.}\ \bibnamefont {Willems~van
  Beveren}}, \bibinfo {author} {\bibfnamefont {B.}~\bibnamefont {Witkamp}},
  \bibinfo {author} {\bibfnamefont {L.~M.~K.}\ \bibnamefont {Vandersypen}},\
  and\ \bibinfo {author} {\bibfnamefont {L.~P.}\ \bibnamefont {Kouwenhoven}},\
  }\href {https://doi.org/10.1038/nature02693} {\bibfield  {journal} {\bibinfo
  {journal} {Nature}\ }\textbf {\bibinfo {volume} {430}},\ \bibinfo {pages}
  {431} (\bibinfo {year} {2004})}\BibitemShut {NoStop}%
\bibitem [{\citenamefont {Petta}\ \emph {et~al.}(2005)\citenamefont {Petta},
  \citenamefont {Johnson}, \citenamefont {Taylor}, \citenamefont {Laird},
  \citenamefont {Yacoby}, \citenamefont {Lukin}, \citenamefont {Marcus},
  \citenamefont {Hanson},\ and\ \citenamefont
  {Gossard}}]{doi:10.1126/science.1116955}%
  \BibitemOpen
  \bibfield  {author} {\bibinfo {author} {\bibfnamefont {J.~R.}\ \bibnamefont
  {Petta}}, \bibinfo {author} {\bibfnamefont {A.~C.}\ \bibnamefont {Johnson}},
  \bibinfo {author} {\bibfnamefont {J.~M.}\ \bibnamefont {Taylor}}, \bibinfo
  {author} {\bibfnamefont {E.~A.}\ \bibnamefont {Laird}}, \bibinfo {author}
  {\bibfnamefont {A.}~\bibnamefont {Yacoby}}, \bibinfo {author} {\bibfnamefont
  {M.~D.}\ \bibnamefont {Lukin}}, \bibinfo {author} {\bibfnamefont {C.~M.}\
  \bibnamefont {Marcus}}, \bibinfo {author} {\bibfnamefont {M.~P.}\
  \bibnamefont {Hanson}},\ and\ \bibinfo {author} {\bibfnamefont {A.~C.}\
  \bibnamefont {Gossard}},\ }\href
  {https://doi.org/doi:10.1126/science.1116955} {\bibfield  {journal} {\bibinfo
   {journal} {Science}\ }\textbf {\bibinfo {volume} {309}},\ \bibinfo {pages}
  {2180} (\bibinfo {year} {2005})}\BibitemShut {NoStop}%
\bibitem [{\citenamefont {Vigneau}\ \emph {et~al.}(2023)\citenamefont
  {Vigneau}, \citenamefont {Fedele}, \citenamefont {Chatterjee}, \citenamefont
  {Reilly}, \citenamefont {Kuemmeth}, \citenamefont {Gonzalez-Zalba},
  \citenamefont {Laird},\ and\ \citenamefont {Ares}}]{2202.10516}%
  \BibitemOpen
  \bibfield  {author} {\bibinfo {author} {\bibfnamefont {F.}~\bibnamefont
  {Vigneau}}, \bibinfo {author} {\bibfnamefont {F.}~\bibnamefont {Fedele}},
  \bibinfo {author} {\bibfnamefont {A.}~\bibnamefont {Chatterjee}}, \bibinfo
  {author} {\bibfnamefont {D.}~\bibnamefont {Reilly}}, \bibinfo {author}
  {\bibfnamefont {F.}~\bibnamefont {Kuemmeth}}, \bibinfo {author}
  {\bibfnamefont {M.~F.}\ \bibnamefont {Gonzalez-Zalba}}, \bibinfo {author}
  {\bibfnamefont {E.}~\bibnamefont {Laird}},\ and\ \bibinfo {author}
  {\bibfnamefont {N.}~\bibnamefont {Ares}},\ }\href
  {https://doi.org/10.1063/5.0088229} {\bibfield  {journal} {\bibinfo
  {journal} {Appl. Phys. Rev.}\ }\textbf {\bibinfo {volume} {10}},\ \bibinfo
  {pages} {021305} (\bibinfo {year} {2023})}\BibitemShut {NoStop}%
\bibitem [{\citenamefont {Reilly}\ \emph {et~al.}(2007)\citenamefont {Reilly},
  \citenamefont {Marcus}, \citenamefont {Hanson},\ and\ \citenamefont
  {Gossard}}]{doi:10.1063/1.2794995}%
  \BibitemOpen
  \bibfield  {author} {\bibinfo {author} {\bibfnamefont {D.~J.}\ \bibnamefont
  {Reilly}}, \bibinfo {author} {\bibfnamefont {C.~M.}\ \bibnamefont {Marcus}},
  \bibinfo {author} {\bibfnamefont {M.~P.}\ \bibnamefont {Hanson}},\ and\
  \bibinfo {author} {\bibfnamefont {A.~C.}\ \bibnamefont {Gossard}},\ }\href
  {https://doi.org/10.1063/1.2794995} {\bibfield  {journal} {\bibinfo
  {journal} {Appl. Phys. Lett.}\ }\textbf {\bibinfo {volume} {91}},\ \bibinfo
  {pages} {162101} (\bibinfo {year} {2007})}\BibitemShut {NoStop}%
\bibitem [{\citenamefont {Blais}\ \emph {et~al.}(2021)\citenamefont {Blais},
  \citenamefont {Grimsmo}, \citenamefont {Girvin},\ and\ \citenamefont
  {Wallraff}}]{RevModPhys.93.025005}%
  \BibitemOpen
  \bibfield  {author} {\bibinfo {author} {\bibfnamefont {A.}~\bibnamefont
  {Blais}}, \bibinfo {author} {\bibfnamefont {A.~L.}\ \bibnamefont {Grimsmo}},
  \bibinfo {author} {\bibfnamefont {S.~M.}\ \bibnamefont {Girvin}},\ and\
  \bibinfo {author} {\bibfnamefont {A.}~\bibnamefont {Wallraff}},\ }\href
  {https://doi.org/10.1103/RevModPhys.93.025005} {\bibfield  {journal}
  {\bibinfo  {journal} {Rev. Mod. Phys.}\ }\textbf {\bibinfo {volume} {93}},\
  \bibinfo {pages} {025005} (\bibinfo {year} {2021})}\BibitemShut {NoStop}%
\bibitem [{\citenamefont {Nielsen}\ and\ \citenamefont
  {Chuang}(2011)}]{nielsen_chuang}%
  \BibitemOpen
  \bibfield  {author} {\bibinfo {author} {\bibfnamefont {M.~A.}\ \bibnamefont
  {Nielsen}}\ and\ \bibinfo {author} {\bibfnamefont {I.~L.}\ \bibnamefont
  {Chuang}},\ }\href@noop {} {\emph {\bibinfo {title} {Quantum Computation and
  Quantum Information: 10th Anniversary Edition}}}\ (\bibinfo  {publisher}
  {Cambridge University Press},\ \bibinfo {year} {2011})\BibitemShut {NoStop}%
\bibitem [{\citenamefont {Fowler}\ \emph {et~al.}(2012)\citenamefont {Fowler},
  \citenamefont {Mariantoni}, \citenamefont {Martinis},\ and\ \citenamefont
  {Cleland}}]{quant-ph/9712048}%
  \BibitemOpen
  \bibfield  {author} {\bibinfo {author} {\bibfnamefont {A.~G.}\ \bibnamefont
  {Fowler}}, \bibinfo {author} {\bibfnamefont {M.}~\bibnamefont {Mariantoni}},
  \bibinfo {author} {\bibfnamefont {J.~M.}\ \bibnamefont {Martinis}},\ and\
  \bibinfo {author} {\bibfnamefont {A.~N.}\ \bibnamefont {Cleland}},\ }\href
  {https://doi.org/10.1103/PhysRevA.86.032324} {\bibfield  {journal} {\bibinfo
  {journal} {Phys. Rev. A}\ }\textbf {\bibinfo {volume} {86}},\ \bibinfo
  {pages} {032324} (\bibinfo {year} {2012})}\BibitemShut {NoStop}%
\bibitem [{\citenamefont {Burkard}\ \emph {et~al.}(ress)\citenamefont
  {Burkard}, \citenamefont {Ladd}, \citenamefont {Nichol}, \citenamefont
  {Pan},\ and\ \citenamefont {Petta}}]{JRP_review}%
  \BibitemOpen
  \bibfield  {author} {\bibinfo {author} {\bibfnamefont {G.}~\bibnamefont
  {Burkard}}, \bibinfo {author} {\bibfnamefont {T.~D.}\ \bibnamefont {Ladd}},
  \bibinfo {author} {\bibfnamefont {J.~M.}\ \bibnamefont {Nichol}}, \bibinfo
  {author} {\bibfnamefont {A.}~\bibnamefont {Pan}},\ and\ \bibinfo {author}
  {\bibfnamefont {J.~R.}\ \bibnamefont {Petta}},\ }\href@noop {} {\bibfield
  {journal} {\bibinfo  {journal} {Rev. Mod. Phys.}\ } (\bibinfo {year} {in
  press})}\BibitemShut {NoStop}%
\bibitem [{\citenamefont {Tyryshkin}\ \emph {et~al.}(2012)\citenamefont
  {Tyryshkin}, \citenamefont {Tojo}, \citenamefont {Morton}, \citenamefont
  {Riemann}, \citenamefont {Abrosimov}, \citenamefont {Becker}, \citenamefont
  {Pohl}, \citenamefont {Schenkel}, \citenamefont {Thewalt}, \citenamefont
  {Itoh},\ and\ \citenamefont {Lyon}}]{Tyryshkin2012}%
  \BibitemOpen
  \bibfield  {author} {\bibinfo {author} {\bibfnamefont {A.~M.}\ \bibnamefont
  {Tyryshkin}}, \bibinfo {author} {\bibfnamefont {S.}~\bibnamefont {Tojo}},
  \bibinfo {author} {\bibfnamefont {J.~J.~L.}\ \bibnamefont {Morton}}, \bibinfo
  {author} {\bibfnamefont {H.}~\bibnamefont {Riemann}}, \bibinfo {author}
  {\bibfnamefont {N.~V.}\ \bibnamefont {Abrosimov}}, \bibinfo {author}
  {\bibfnamefont {P.}~\bibnamefont {Becker}}, \bibinfo {author} {\bibfnamefont
  {H.-J.}\ \bibnamefont {Pohl}}, \bibinfo {author} {\bibfnamefont
  {T.}~\bibnamefont {Schenkel}}, \bibinfo {author} {\bibfnamefont {M.~L.~W.}\
  \bibnamefont {Thewalt}}, \bibinfo {author} {\bibfnamefont {K.~M.}\
  \bibnamefont {Itoh}},\ and\ \bibinfo {author} {\bibfnamefont {S.~A.}\
  \bibnamefont {Lyon}},\ }\href {https://doi.org/10.1038/nmat3182} {\bibfield
  {journal} {\bibinfo  {journal} {Nat. Mater.}\ }\textbf {\bibinfo {volume}
  {11}},\ \bibinfo {pages} {143} (\bibinfo {year} {2012})}\BibitemShut
  {NoStop}%
\bibitem [{\citenamefont {Mills}\ \emph {et~al.}(2022)\citenamefont {Mills},
  \citenamefont {Guinn}, \citenamefont {Gullans}, \citenamefont {Sigillito},
  \citenamefont {Feldman}, \citenamefont {Nielsen},\ and\ \citenamefont
  {Petta}}]{doi:10.1126/sciadv.abn5130}%
  \BibitemOpen
  \bibfield  {author} {\bibinfo {author} {\bibfnamefont {A.~R.}\ \bibnamefont
  {Mills}}, \bibinfo {author} {\bibfnamefont {C.~R.}\ \bibnamefont {Guinn}},
  \bibinfo {author} {\bibfnamefont {M.~J.}\ \bibnamefont {Gullans}}, \bibinfo
  {author} {\bibfnamefont {A.~J.}\ \bibnamefont {Sigillito}}, \bibinfo {author}
  {\bibfnamefont {M.~M.}\ \bibnamefont {Feldman}}, \bibinfo {author}
  {\bibfnamefont {E.}~\bibnamefont {Nielsen}},\ and\ \bibinfo {author}
  {\bibfnamefont {J.~R.}\ \bibnamefont {Petta}},\ }\href
  {https://doi.org/10.1126/sciadv.abn5130} {\bibfield  {journal} {\bibinfo
  {journal} {Sci. Adv.}\ }\textbf {\bibinfo {volume} {8}},\ \bibinfo {pages}
  {eabn5130} (\bibinfo {year} {2022})}\BibitemShut {NoStop}%
\bibitem [{\citenamefont {Xue}\ \emph {et~al.}(2022)\citenamefont {Xue},
  \citenamefont {Russ}, \citenamefont {Samkharadze}, \citenamefont {Undseth},
  \citenamefont {Sammak}, \citenamefont {Scappucci},\ and\ \citenamefont
  {Vandersypen}}]{Xue2022}%
  \BibitemOpen
  \bibfield  {author} {\bibinfo {author} {\bibfnamefont {X.}~\bibnamefont
  {Xue}}, \bibinfo {author} {\bibfnamefont {M.}~\bibnamefont {Russ}}, \bibinfo
  {author} {\bibfnamefont {N.}~\bibnamefont {Samkharadze}}, \bibinfo {author}
  {\bibfnamefont {B.}~\bibnamefont {Undseth}}, \bibinfo {author} {\bibfnamefont
  {A.}~\bibnamefont {Sammak}}, \bibinfo {author} {\bibfnamefont
  {G.}~\bibnamefont {Scappucci}},\ and\ \bibinfo {author} {\bibfnamefont
  {L.~M.~K.}\ \bibnamefont {Vandersypen}},\ }\href
  {https://doi.org/10.1038/s41586-021-04273-w} {\bibfield  {journal} {\bibinfo
  {journal} {Nature}\ }\textbf {\bibinfo {volume} {601}},\ \bibinfo {pages}
  {343} (\bibinfo {year} {2022})}\BibitemShut {NoStop}%
\bibitem [{\citenamefont {Noiri}\ \emph {et~al.}(2022)\citenamefont {Noiri},
  \citenamefont {Takeda}, \citenamefont {Nakajima}, \citenamefont {Kobayashi},
  \citenamefont {Sammak}, \citenamefont {Scappucci},\ and\ \citenamefont
  {Tarucha}}]{Noiri2022}%
  \BibitemOpen
  \bibfield  {author} {\bibinfo {author} {\bibfnamefont {A.}~\bibnamefont
  {Noiri}}, \bibinfo {author} {\bibfnamefont {K.}~\bibnamefont {Takeda}},
  \bibinfo {author} {\bibfnamefont {T.}~\bibnamefont {Nakajima}}, \bibinfo
  {author} {\bibfnamefont {T.}~\bibnamefont {Kobayashi}}, \bibinfo {author}
  {\bibfnamefont {A.}~\bibnamefont {Sammak}}, \bibinfo {author} {\bibfnamefont
  {G.}~\bibnamefont {Scappucci}},\ and\ \bibinfo {author} {\bibfnamefont
  {S.}~\bibnamefont {Tarucha}},\ }\href
  {https://doi.org/10.1038/s41586-021-04182-y} {\bibfield  {journal} {\bibinfo
  {journal} {Nature}\ }\textbf {\bibinfo {volume} {601}},\ \bibinfo {pages}
  {338} (\bibinfo {year} {2022})}\BibitemShut {NoStop}%
\bibitem [{\citenamefont {Maurand}\ \emph {et~al.}(2016)\citenamefont
  {Maurand}, \citenamefont {Jehl}, \citenamefont {Kotekar-Patil}, \citenamefont
  {Corna}, \citenamefont {Bohuslavskyi}, \citenamefont {Lavi{\'e}ville},
  \citenamefont {Hutin}, \citenamefont {Barraud}, \citenamefont {Vinet},
  \citenamefont {Sanquer},\ and\ \citenamefont {De~Franceschi}}]{Maurand2016}%
  \BibitemOpen
  \bibfield  {author} {\bibinfo {author} {\bibfnamefont {R.}~\bibnamefont
  {Maurand}}, \bibinfo {author} {\bibfnamefont {X.}~\bibnamefont {Jehl}},
  \bibinfo {author} {\bibfnamefont {D.}~\bibnamefont {Kotekar-Patil}}, \bibinfo
  {author} {\bibfnamefont {A.}~\bibnamefont {Corna}}, \bibinfo {author}
  {\bibfnamefont {H.}~\bibnamefont {Bohuslavskyi}}, \bibinfo {author}
  {\bibfnamefont {R.}~\bibnamefont {Lavi{\'e}ville}}, \bibinfo {author}
  {\bibfnamefont {L.}~\bibnamefont {Hutin}}, \bibinfo {author} {\bibfnamefont
  {S.}~\bibnamefont {Barraud}}, \bibinfo {author} {\bibfnamefont
  {M.}~\bibnamefont {Vinet}}, \bibinfo {author} {\bibfnamefont
  {M.}~\bibnamefont {Sanquer}},\ and\ \bibinfo {author} {\bibfnamefont
  {S.}~\bibnamefont {De~Franceschi}},\ }\href
  {https://doi.org/10.1038/ncomms13575} {\bibfield  {journal} {\bibinfo
  {journal} {Nat. Commun.}\ }\textbf {\bibinfo {volume} {7}},\ \bibinfo {pages}
  {13575} (\bibinfo {year} {2016})}\BibitemShut {NoStop}%
\bibitem [{\citenamefont {Zwerver}\ \emph {et~al.}(2022)\citenamefont
  {Zwerver}, \citenamefont {Kr{\"a}henmann}, \citenamefont {Watson},
  \citenamefont {Lampert}, \citenamefont {George}, \citenamefont
  {Pillarisetty}, \citenamefont {Bojarski}, \citenamefont {Amin}, \citenamefont
  {Amitonov}, \citenamefont {Boter}, \citenamefont {Caudillo}, \citenamefont
  {Correas-Serrano}, \citenamefont {Dehollain}, \citenamefont {Droulers},
  \citenamefont {Henry}, \citenamefont {Kotlyar}, \citenamefont {Lodari},
  \citenamefont {L{\"u}thi}, \citenamefont {Michalak}, \citenamefont {Mueller},
  \citenamefont {Neyens}, \citenamefont {Roberts}, \citenamefont {Samkharadze},
  \citenamefont {Zheng}, \citenamefont {Zietz}, \citenamefont {Scappucci},
  \citenamefont {Veldhorst}, \citenamefont {Vandersypen},\ and\ \citenamefont
  {Clarke}}]{Zwerver2022}%
  \BibitemOpen
  \bibfield  {author} {\bibinfo {author} {\bibfnamefont {A.~M.~J.}\
  \bibnamefont {Zwerver}}, \bibinfo {author} {\bibfnamefont {T.}~\bibnamefont
  {Kr{\"a}henmann}}, \bibinfo {author} {\bibfnamefont {T.~F.}\ \bibnamefont
  {Watson}}, \bibinfo {author} {\bibfnamefont {L.}~\bibnamefont {Lampert}},
  \bibinfo {author} {\bibfnamefont {H.~C.}\ \bibnamefont {George}}, \bibinfo
  {author} {\bibfnamefont {R.}~\bibnamefont {Pillarisetty}}, \bibinfo {author}
  {\bibfnamefont {S.~A.}\ \bibnamefont {Bojarski}}, \bibinfo {author}
  {\bibfnamefont {P.}~\bibnamefont {Amin}}, \bibinfo {author} {\bibfnamefont
  {S.~V.}\ \bibnamefont {Amitonov}}, \bibinfo {author} {\bibfnamefont {J.~M.}\
  \bibnamefont {Boter}}, \bibinfo {author} {\bibfnamefont {R.}~\bibnamefont
  {Caudillo}}, \bibinfo {author} {\bibfnamefont {D.}~\bibnamefont
  {Correas-Serrano}}, \bibinfo {author} {\bibfnamefont {J.~P.}\ \bibnamefont
  {Dehollain}}, \bibinfo {author} {\bibfnamefont {G.}~\bibnamefont {Droulers}},
  \bibinfo {author} {\bibfnamefont {E.~M.}\ \bibnamefont {Henry}}, \bibinfo
  {author} {\bibfnamefont {R.}~\bibnamefont {Kotlyar}}, \bibinfo {author}
  {\bibfnamefont {M.}~\bibnamefont {Lodari}}, \bibinfo {author} {\bibfnamefont
  {F.}~\bibnamefont {L{\"u}thi}}, \bibinfo {author} {\bibfnamefont {D.~J.}\
  \bibnamefont {Michalak}}, \bibinfo {author} {\bibfnamefont {B.~K.}\
  \bibnamefont {Mueller}}, \bibinfo {author} {\bibfnamefont {S.}~\bibnamefont
  {Neyens}}, \bibinfo {author} {\bibfnamefont {J.}~\bibnamefont {Roberts}},
  \bibinfo {author} {\bibfnamefont {N.}~\bibnamefont {Samkharadze}}, \bibinfo
  {author} {\bibfnamefont {G.}~\bibnamefont {Zheng}}, \bibinfo {author}
  {\bibfnamefont {O.~K.}\ \bibnamefont {Zietz}}, \bibinfo {author}
  {\bibfnamefont {G.}~\bibnamefont {Scappucci}}, \bibinfo {author}
  {\bibfnamefont {M.}~\bibnamefont {Veldhorst}}, \bibinfo {author}
  {\bibfnamefont {L.~M.~K.}\ \bibnamefont {Vandersypen}},\ and\ \bibinfo
  {author} {\bibfnamefont {J.~S.}\ \bibnamefont {Clarke}},\ }\href
  {https://doi.org/10.1038/s41928-022-00727-9} {\bibfield  {journal} {\bibinfo
  {journal} {Nat. Electron.}\ }\textbf {\bibinfo {volume} {5}},\ \bibinfo
  {pages} {184} (\bibinfo {year} {2022})}\BibitemShut {NoStop}%
\bibitem [{\citenamefont {Xue}\ \emph {et~al.}(2021)\citenamefont {Xue},
  \citenamefont {Patra}, \citenamefont {van Dijk}, \citenamefont {Samkharadze},
  \citenamefont {Subramanian}, \citenamefont {Corna}, \citenamefont
  {Paquelet~Wuetz}, \citenamefont {Jeon}, \citenamefont {Sheikh}, \citenamefont
  {Juarez-Hernandez}, \citenamefont {Esparza}, \citenamefont {Rampurawala},
  \citenamefont {Carlton}, \citenamefont {Ravikumar}, \citenamefont {Nieva},
  \citenamefont {Kim}, \citenamefont {Lee}, \citenamefont {Sammak},
  \citenamefont {Scappucci}, \citenamefont {Veldhorst}, \citenamefont
  {Sebastiano}, \citenamefont {Babaie}, \citenamefont {Pellerano},
  \citenamefont {Charbon},\ and\ \citenamefont {Vandersypen}}]{Xue2021}%
  \BibitemOpen
  \bibfield  {author} {\bibinfo {author} {\bibfnamefont {X.}~\bibnamefont
  {Xue}}, \bibinfo {author} {\bibfnamefont {B.}~\bibnamefont {Patra}}, \bibinfo
  {author} {\bibfnamefont {J.~P.~G.}\ \bibnamefont {van Dijk}}, \bibinfo
  {author} {\bibfnamefont {N.}~\bibnamefont {Samkharadze}}, \bibinfo {author}
  {\bibfnamefont {S.}~\bibnamefont {Subramanian}}, \bibinfo {author}
  {\bibfnamefont {A.}~\bibnamefont {Corna}}, \bibinfo {author} {\bibfnamefont
  {B.}~\bibnamefont {Paquelet~Wuetz}}, \bibinfo {author} {\bibfnamefont
  {C.}~\bibnamefont {Jeon}}, \bibinfo {author} {\bibfnamefont {F.}~\bibnamefont
  {Sheikh}}, \bibinfo {author} {\bibfnamefont {E.}~\bibnamefont
  {Juarez-Hernandez}}, \bibinfo {author} {\bibfnamefont {B.~P.}\ \bibnamefont
  {Esparza}}, \bibinfo {author} {\bibfnamefont {H.}~\bibnamefont
  {Rampurawala}}, \bibinfo {author} {\bibfnamefont {B.}~\bibnamefont
  {Carlton}}, \bibinfo {author} {\bibfnamefont {S.}~\bibnamefont {Ravikumar}},
  \bibinfo {author} {\bibfnamefont {C.}~\bibnamefont {Nieva}}, \bibinfo
  {author} {\bibfnamefont {S.}~\bibnamefont {Kim}}, \bibinfo {author}
  {\bibfnamefont {H.-J.}\ \bibnamefont {Lee}}, \bibinfo {author} {\bibfnamefont
  {A.}~\bibnamefont {Sammak}}, \bibinfo {author} {\bibfnamefont
  {G.}~\bibnamefont {Scappucci}}, \bibinfo {author} {\bibfnamefont
  {M.}~\bibnamefont {Veldhorst}}, \bibinfo {author} {\bibfnamefont
  {F.}~\bibnamefont {Sebastiano}}, \bibinfo {author} {\bibfnamefont
  {M.}~\bibnamefont {Babaie}}, \bibinfo {author} {\bibfnamefont
  {S.}~\bibnamefont {Pellerano}}, \bibinfo {author} {\bibfnamefont
  {E.}~\bibnamefont {Charbon}},\ and\ \bibinfo {author} {\bibfnamefont
  {L.~M.~K.}\ \bibnamefont {Vandersypen}},\ }\href
  {https://doi.org/10.1038/s41586-021-03469-4} {\bibfield  {journal} {\bibinfo
  {journal} {Nature}\ }\textbf {\bibinfo {volume} {593}},\ \bibinfo {pages}
  {205} (\bibinfo {year} {2021})}\BibitemShut {NoStop}%
\bibitem [{\citenamefont {Barthel}\ \emph {et~al.}(2009)\citenamefont
  {Barthel}, \citenamefont {Reilly}, \citenamefont {Marcus}, \citenamefont
  {Hanson},\ and\ \citenamefont {Gossard}}]{PhysRevLett.103.160503}%
  \BibitemOpen
  \bibfield  {author} {\bibinfo {author} {\bibfnamefont {C.}~\bibnamefont
  {Barthel}}, \bibinfo {author} {\bibfnamefont {D.~J.}\ \bibnamefont {Reilly}},
  \bibinfo {author} {\bibfnamefont {C.~M.}\ \bibnamefont {Marcus}}, \bibinfo
  {author} {\bibfnamefont {M.~P.}\ \bibnamefont {Hanson}},\ and\ \bibinfo
  {author} {\bibfnamefont {A.~C.}\ \bibnamefont {Gossard}},\ }\href
  {https://doi.org/10.1103/PhysRevLett.103.160503} {\bibfield  {journal}
  {\bibinfo  {journal} {Phys. Rev. Lett.}\ }\textbf {\bibinfo {volume} {103}},\
  \bibinfo {pages} {160503} (\bibinfo {year} {2009})}\BibitemShut {NoStop}%
\bibitem [{\citenamefont {Hanson}\ \emph {et~al.}(2007)\citenamefont {Hanson},
  \citenamefont {Kouwenhoven}, \citenamefont {Petta}, \citenamefont {Tarucha},\
  and\ \citenamefont {Vandersypen}}]{RevModPhys.79.1217}%
  \BibitemOpen
  \bibfield  {author} {\bibinfo {author} {\bibfnamefont {R.}~\bibnamefont
  {Hanson}}, \bibinfo {author} {\bibfnamefont {L.~P.}\ \bibnamefont
  {Kouwenhoven}}, \bibinfo {author} {\bibfnamefont {J.~R.}\ \bibnamefont
  {Petta}}, \bibinfo {author} {\bibfnamefont {S.}~\bibnamefont {Tarucha}},\
  and\ \bibinfo {author} {\bibfnamefont {L.~M.~K.}\ \bibnamefont
  {Vandersypen}},\ }\href {https://doi.org/10.1103/RevModPhys.79.1217}
  {\bibfield  {journal} {\bibinfo  {journal} {Rev. Mod. Phys.}\ }\textbf
  {\bibinfo {volume} {79}},\ \bibinfo {pages} {1217} (\bibinfo {year}
  {2007})}\BibitemShut {NoStop}%
\bibitem [{\citenamefont {Schoelkopf}\ \emph {et~al.}(1998)\citenamefont
  {Schoelkopf}, \citenamefont {Wahlgren}, \citenamefont {Kozhevnikov},
  \citenamefont {Delsing},\ and\ \citenamefont
  {Prober}}]{doi:10.1126/science.280.5367.1238}%
  \BibitemOpen
  \bibfield  {author} {\bibinfo {author} {\bibfnamefont {R.~J.}\ \bibnamefont
  {Schoelkopf}}, \bibinfo {author} {\bibfnamefont {P.}~\bibnamefont
  {Wahlgren}}, \bibinfo {author} {\bibfnamefont {A.~A.}\ \bibnamefont
  {Kozhevnikov}}, \bibinfo {author} {\bibfnamefont {P.}~\bibnamefont
  {Delsing}},\ and\ \bibinfo {author} {\bibfnamefont {D.~E.}\ \bibnamefont
  {Prober}},\ }\href {https://doi.org/10.1126/science.280.5367.1238} {\bibfield
   {journal} {\bibinfo  {journal} {Science}\ }\textbf {\bibinfo {volume}
  {280}},\ \bibinfo {pages} {1238} (\bibinfo {year} {1998})}\BibitemShut
  {NoStop}%
\bibitem [{\citenamefont {Colless}\ \emph {et~al.}(2013)\citenamefont
  {Colless}, \citenamefont {Mahoney}, \citenamefont {Hornibrook}, \citenamefont
  {Doherty}, \citenamefont {Lu}, \citenamefont {Gossard},\ and\ \citenamefont
  {Reilly}}]{PhysRevLett.110.046805}%
  \BibitemOpen
  \bibfield  {author} {\bibinfo {author} {\bibfnamefont {J.~I.}\ \bibnamefont
  {Colless}}, \bibinfo {author} {\bibfnamefont {A.~C.}\ \bibnamefont
  {Mahoney}}, \bibinfo {author} {\bibfnamefont {J.~M.}\ \bibnamefont
  {Hornibrook}}, \bibinfo {author} {\bibfnamefont {A.~C.}\ \bibnamefont
  {Doherty}}, \bibinfo {author} {\bibfnamefont {H.}~\bibnamefont {Lu}},
  \bibinfo {author} {\bibfnamefont {A.~C.}\ \bibnamefont {Gossard}},\ and\
  \bibinfo {author} {\bibfnamefont {D.~J.}\ \bibnamefont {Reilly}},\ }\href
  {https://doi.org/10.1103/PhysRevLett.110.046805} {\bibfield  {journal}
  {\bibinfo  {journal} {Phys. Rev. Lett.}\ }\textbf {\bibinfo {volume} {110}},\
  \bibinfo {pages} {046805} (\bibinfo {year} {2013})}\BibitemShut {NoStop}%
\bibitem [{\citenamefont {Crippa}\ \emph {et~al.}(2019)\citenamefont {Crippa},
  \citenamefont {Ezzouch}, \citenamefont {Apr{\'a}}, \citenamefont {Amisse},
  \citenamefont {Lavi{\'e}ville}, \citenamefont {Hutin}, \citenamefont
  {Bertrand}, \citenamefont {Vinet}, \citenamefont {Urdampilleta},
  \citenamefont {Meunier}, \citenamefont {Sanquer}, \citenamefont {Jehl},
  \citenamefont {Maurand},\ and\ \citenamefont {De~Franceschi}}]{Crippa2019}%
  \BibitemOpen
  \bibfield  {author} {\bibinfo {author} {\bibfnamefont {A.}~\bibnamefont
  {Crippa}}, \bibinfo {author} {\bibfnamefont {R.}~\bibnamefont {Ezzouch}},
  \bibinfo {author} {\bibfnamefont {A.}~\bibnamefont {Apr{\'a}}}, \bibinfo
  {author} {\bibfnamefont {A.}~\bibnamefont {Amisse}}, \bibinfo {author}
  {\bibfnamefont {R.}~\bibnamefont {Lavi{\'e}ville}}, \bibinfo {author}
  {\bibfnamefont {L.}~\bibnamefont {Hutin}}, \bibinfo {author} {\bibfnamefont
  {B.}~\bibnamefont {Bertrand}}, \bibinfo {author} {\bibfnamefont
  {M.}~\bibnamefont {Vinet}}, \bibinfo {author} {\bibfnamefont
  {M.}~\bibnamefont {Urdampilleta}}, \bibinfo {author} {\bibfnamefont
  {T.}~\bibnamefont {Meunier}}, \bibinfo {author} {\bibfnamefont
  {M.}~\bibnamefont {Sanquer}}, \bibinfo {author} {\bibfnamefont
  {X.}~\bibnamefont {Jehl}}, \bibinfo {author} {\bibfnamefont {R.}~\bibnamefont
  {Maurand}},\ and\ \bibinfo {author} {\bibfnamefont {S.}~\bibnamefont
  {De~Franceschi}},\ }\href {https://doi.org/10.1038/s41467-019-10848-z}
  {\bibfield  {journal} {\bibinfo  {journal} {Nat. Commun.}\ }\textbf {\bibinfo
  {volume} {10}},\ \bibinfo {pages} {2776} (\bibinfo {year}
  {2019})}\BibitemShut {NoStop}%
\bibitem [{\citenamefont {Gonzalez-Zalba}\ \emph {et~al.}(2016)\citenamefont
  {Gonzalez-Zalba}, \citenamefont {Shevchenko}, \citenamefont {Barraud},
  \citenamefont {Johansson}, \citenamefont {Ferguson}, \citenamefont {Nori},\
  and\ \citenamefont {Betz}}]{doi:10.1021/acs.nanolett.5b04356}%
  \BibitemOpen
  \bibfield  {author} {\bibinfo {author} {\bibfnamefont {M.~F.}\ \bibnamefont
  {Gonzalez-Zalba}}, \bibinfo {author} {\bibfnamefont {S.~N.}\ \bibnamefont
  {Shevchenko}}, \bibinfo {author} {\bibfnamefont {S.}~\bibnamefont {Barraud}},
  \bibinfo {author} {\bibfnamefont {J.~R.}\ \bibnamefont {Johansson}}, \bibinfo
  {author} {\bibfnamefont {A.~J.}\ \bibnamefont {Ferguson}}, \bibinfo {author}
  {\bibfnamefont {F.}~\bibnamefont {Nori}},\ and\ \bibinfo {author}
  {\bibfnamefont {A.~C.}\ \bibnamefont {Betz}},\ }\href
  {https://doi.org/10.1021/acs.nanolett.5b04356} {\bibfield  {journal}
  {\bibinfo  {journal} {Nano Lett.}\ }\textbf {\bibinfo {volume} {16}},\
  \bibinfo {pages} {1614} (\bibinfo {year} {2016})}\BibitemShut {NoStop}%
\bibitem [{\citenamefont {Betz}\ \emph {et~al.}(2015)\citenamefont {Betz},
  \citenamefont {Wacquez}, \citenamefont {Vinet}, \citenamefont {Jehl},
  \citenamefont {Saraiva}, \citenamefont {Sanquer}, \citenamefont {Ferguson},\
  and\ \citenamefont {Gonzalez-Zalba}}]{doi:10.1021/acs.nanolett.5b01306}%
  \BibitemOpen
  \bibfield  {author} {\bibinfo {author} {\bibfnamefont {A.~C.}\ \bibnamefont
  {Betz}}, \bibinfo {author} {\bibfnamefont {R.}~\bibnamefont {Wacquez}},
  \bibinfo {author} {\bibfnamefont {M.}~\bibnamefont {Vinet}}, \bibinfo
  {author} {\bibfnamefont {X.}~\bibnamefont {Jehl}}, \bibinfo {author}
  {\bibfnamefont {A.~L.}\ \bibnamefont {Saraiva}}, \bibinfo {author}
  {\bibfnamefont {M.}~\bibnamefont {Sanquer}}, \bibinfo {author} {\bibfnamefont
  {A.~J.}\ \bibnamefont {Ferguson}},\ and\ \bibinfo {author} {\bibfnamefont
  {M.~F.}\ \bibnamefont {Gonzalez-Zalba}},\ }\href
  {https://doi.org/10.1021/acs.nanolett.5b01306} {\bibfield  {journal}
  {\bibinfo  {journal} {Nano Lett.}\ }\textbf {\bibinfo {volume} {15}},\
  \bibinfo {pages} {4622} (\bibinfo {year} {2015})}\BibitemShut {NoStop}%
\bibitem [{\citenamefont {Rossi}\ \emph {et~al.}(2017)\citenamefont {Rossi},
  \citenamefont {Zhao}, \citenamefont {Dzurak},\ and\ \citenamefont
  {Gonzalez-Zalba}}]{Rossi}%
  \BibitemOpen
  \bibfield  {author} {\bibinfo {author} {\bibfnamefont {A.}~\bibnamefont
  {Rossi}}, \bibinfo {author} {\bibfnamefont {R.}~\bibnamefont {Zhao}},
  \bibinfo {author} {\bibfnamefont {A.~S.}\ \bibnamefont {Dzurak}},\ and\
  \bibinfo {author} {\bibfnamefont {M.~F.}\ \bibnamefont {Gonzalez-Zalba}},\
  }\href {https://doi.org/10.1063/1.4984224} {\bibfield  {journal} {\bibinfo
  {journal} {Appl. Phys. Lett.}\ }\textbf {\bibinfo {volume} {110}},\ \bibinfo
  {pages} {212101} (\bibinfo {year} {2017})}\BibitemShut {NoStop}%
\bibitem [{\citenamefont {West}\ \emph {et~al.}(2019)\citenamefont {West},
  \citenamefont {Hensen}, \citenamefont {Jouan}, \citenamefont {Tanttu},
  \citenamefont {Yang}, \citenamefont {Rossi}, \citenamefont {Gonzalez-Zalba},
  \citenamefont {Hudson}, \citenamefont {Morello}, \citenamefont {Reilly},\
  and\ \citenamefont {Dzurak}}]{West2019}%
  \BibitemOpen
  \bibfield  {author} {\bibinfo {author} {\bibfnamefont {A.}~\bibnamefont
  {West}}, \bibinfo {author} {\bibfnamefont {B.}~\bibnamefont {Hensen}},
  \bibinfo {author} {\bibfnamefont {A.}~\bibnamefont {Jouan}}, \bibinfo
  {author} {\bibfnamefont {T.}~\bibnamefont {Tanttu}}, \bibinfo {author}
  {\bibfnamefont {C.-H.}\ \bibnamefont {Yang}}, \bibinfo {author}
  {\bibfnamefont {A.}~\bibnamefont {Rossi}}, \bibinfo {author} {\bibfnamefont
  {M.~F.}\ \bibnamefont {Gonzalez-Zalba}}, \bibinfo {author} {\bibfnamefont
  {F.}~\bibnamefont {Hudson}}, \bibinfo {author} {\bibfnamefont
  {A.}~\bibnamefont {Morello}}, \bibinfo {author} {\bibfnamefont {D.~J.}\
  \bibnamefont {Reilly}},\ and\ \bibinfo {author} {\bibfnamefont {A.~S.}\
  \bibnamefont {Dzurak}},\ }\href {https://doi.org/10.1038/s41565-019-0400-7}
  {\bibfield  {journal} {\bibinfo  {journal} {Nat. Nanotechnol.}\ }\textbf
  {\bibinfo {volume} {14}},\ \bibinfo {pages} {437} (\bibinfo {year}
  {2019})}\BibitemShut {NoStop}%
\bibitem [{\citenamefont {Zheng}\ \emph {et~al.}(2019)\citenamefont {Zheng},
  \citenamefont {Samkharadze}, \citenamefont {Noordam}, \citenamefont {Kalhor},
  \citenamefont {Brousse}, \citenamefont {Sammak}, \citenamefont {Scappucci},\
  and\ \citenamefont {Vandersypen}}]{Zheng2019}%
  \BibitemOpen
  \bibfield  {author} {\bibinfo {author} {\bibfnamefont {G.}~\bibnamefont
  {Zheng}}, \bibinfo {author} {\bibfnamefont {N.}~\bibnamefont {Samkharadze}},
  \bibinfo {author} {\bibfnamefont {M.~L.}\ \bibnamefont {Noordam}}, \bibinfo
  {author} {\bibfnamefont {N.}~\bibnamefont {Kalhor}}, \bibinfo {author}
  {\bibfnamefont {D.}~\bibnamefont {Brousse}}, \bibinfo {author} {\bibfnamefont
  {A.}~\bibnamefont {Sammak}}, \bibinfo {author} {\bibfnamefont
  {G.}~\bibnamefont {Scappucci}},\ and\ \bibinfo {author} {\bibfnamefont
  {L.~M.~K.}\ \bibnamefont {Vandersypen}},\ }\href
  {https://doi.org/10.1038/s41565-019-0488-9} {\bibfield  {journal} {\bibinfo
  {journal} {Nat. Nanotechnol.}\ }\textbf {\bibinfo {volume} {14}},\ \bibinfo
  {pages} {742} (\bibinfo {year} {2019})}\BibitemShut {NoStop}%
\bibitem [{\citenamefont {Pakkiam}\ \emph {et~al.}(2018)\citenamefont
  {Pakkiam}, \citenamefont {Timofeev}, \citenamefont {House}, \citenamefont
  {Hogg}, \citenamefont {Kobayashi}, \citenamefont {Koch}, \citenamefont
  {Rogge},\ and\ \citenamefont {Simmons}}]{PhysRevX.8.041032}%
  \BibitemOpen
  \bibfield  {author} {\bibinfo {author} {\bibfnamefont {P.}~\bibnamefont
  {Pakkiam}}, \bibinfo {author} {\bibfnamefont {A.~V.}\ \bibnamefont
  {Timofeev}}, \bibinfo {author} {\bibfnamefont {M.~G.}\ \bibnamefont {House}},
  \bibinfo {author} {\bibfnamefont {M.~R.}\ \bibnamefont {Hogg}}, \bibinfo
  {author} {\bibfnamefont {T.}~\bibnamefont {Kobayashi}}, \bibinfo {author}
  {\bibfnamefont {M.}~\bibnamefont {Koch}}, \bibinfo {author} {\bibfnamefont
  {S.}~\bibnamefont {Rogge}},\ and\ \bibinfo {author} {\bibfnamefont {M.~Y.}\
  \bibnamefont {Simmons}},\ }\href {https://doi.org/10.1103/PhysRevX.8.041032}
  {\bibfield  {journal} {\bibinfo  {journal} {Phys. Rev. X}\ }\textbf {\bibinfo
  {volume} {8}},\ \bibinfo {pages} {041032} (\bibinfo {year}
  {2018})}\BibitemShut {NoStop}%
\bibitem [{\citenamefont {Urdampilleta}\ \emph {et~al.}(2019)\citenamefont
  {Urdampilleta}, \citenamefont {Niegemann}, \citenamefont {Chanrion},
  \citenamefont {Jadot}, \citenamefont {Spence}, \citenamefont {Mortemousque},
  \citenamefont {B{\"a}uerle}, \citenamefont {Hutin}, \citenamefont {Bertrand},
  \citenamefont {Barraud}, \citenamefont {Maurand}, \citenamefont {Sanquer},
  \citenamefont {Jehl}, \citenamefont {De~Franceschi}, \citenamefont {Vinet},\
  and\ \citenamefont {Meunier}}]{Urdampilleta2019}%
  \BibitemOpen
  \bibfield  {author} {\bibinfo {author} {\bibfnamefont {M.}~\bibnamefont
  {Urdampilleta}}, \bibinfo {author} {\bibfnamefont {D.~J.}\ \bibnamefont
  {Niegemann}}, \bibinfo {author} {\bibfnamefont {E.}~\bibnamefont {Chanrion}},
  \bibinfo {author} {\bibfnamefont {B.}~\bibnamefont {Jadot}}, \bibinfo
  {author} {\bibfnamefont {C.}~\bibnamefont {Spence}}, \bibinfo {author}
  {\bibfnamefont {P.-A.}\ \bibnamefont {Mortemousque}}, \bibinfo {author}
  {\bibfnamefont {C.}~\bibnamefont {B{\"a}uerle}}, \bibinfo {author}
  {\bibfnamefont {L.}~\bibnamefont {Hutin}}, \bibinfo {author} {\bibfnamefont
  {B.}~\bibnamefont {Bertrand}}, \bibinfo {author} {\bibfnamefont
  {S.}~\bibnamefont {Barraud}}, \bibinfo {author} {\bibfnamefont
  {R.}~\bibnamefont {Maurand}}, \bibinfo {author} {\bibfnamefont
  {M.}~\bibnamefont {Sanquer}}, \bibinfo {author} {\bibfnamefont
  {X.}~\bibnamefont {Jehl}}, \bibinfo {author} {\bibfnamefont {S.}~\bibnamefont
  {De~Franceschi}}, \bibinfo {author} {\bibfnamefont {M.}~\bibnamefont
  {Vinet}},\ and\ \bibinfo {author} {\bibfnamefont {T.}~\bibnamefont
  {Meunier}},\ }\href {https://doi.org/10.1038/s41565-019-0443-9} {\bibfield
  {journal} {\bibinfo  {journal} {Nat. Nanotechnol.}\ }\textbf {\bibinfo
  {volume} {14}},\ \bibinfo {pages} {737} (\bibinfo {year} {2019})}\BibitemShut
  {NoStop}%
\bibitem [{\citenamefont {Oh}\ \emph {et~al.}(2021)\citenamefont {Oh},
  \citenamefont {Denisov}, \citenamefont {Chen},\ and\ \citenamefont
  {Petta}}]{Seong_AIP}%
  \BibitemOpen
  \bibfield  {author} {\bibinfo {author} {\bibfnamefont {S.~W.}\ \bibnamefont
  {Oh}}, \bibinfo {author} {\bibfnamefont {A.~O.}\ \bibnamefont {Denisov}},
  \bibinfo {author} {\bibfnamefont {P.}~\bibnamefont {Chen}},\ and\ \bibinfo
  {author} {\bibfnamefont {J.~R.}\ \bibnamefont {Petta}},\ }\href
  {https://doi.org/10.1063/5.0056648} {\bibfield  {journal} {\bibinfo
  {journal} {AIP Adv.}\ }\textbf {\bibinfo {volume} {11}},\ \bibinfo {pages}
  {125122} (\bibinfo {year} {2021})}\BibitemShut {NoStop}%
\bibitem [{\citenamefont {Denisov}\ \emph {et~al.}(2022)\citenamefont
  {Denisov}, \citenamefont {Oh}, \citenamefont {Fuchs}, \citenamefont {Mills},
  \citenamefont {Chen}, \citenamefont {Anderson}, \citenamefont {Gyure},
  \citenamefont {Barnard},\ and\ \citenamefont
  {Petta}}]{doi:10.1021/acs.nanolett.2c01098}%
  \BibitemOpen
  \bibfield  {author} {\bibinfo {author} {\bibfnamefont {A.~O.}\ \bibnamefont
  {Denisov}}, \bibinfo {author} {\bibfnamefont {S.~W.}\ \bibnamefont {Oh}},
  \bibinfo {author} {\bibfnamefont {G.}~\bibnamefont {Fuchs}}, \bibinfo
  {author} {\bibfnamefont {A.~R.}\ \bibnamefont {Mills}}, \bibinfo {author}
  {\bibfnamefont {P.}~\bibnamefont {Chen}}, \bibinfo {author} {\bibfnamefont
  {C.~R.}\ \bibnamefont {Anderson}}, \bibinfo {author} {\bibfnamefont {M.~F.}\
  \bibnamefont {Gyure}}, \bibinfo {author} {\bibfnamefont {A.~W.}\ \bibnamefont
  {Barnard}},\ and\ \bibinfo {author} {\bibfnamefont {J.~R.}\ \bibnamefont
  {Petta}},\ }\href {https://doi.org/10.1021/acs.nanolett.2c01098} {\bibfield
  {journal} {\bibinfo  {journal} {Nano Lett.}\ }\textbf {\bibinfo {volume}
  {22}},\ \bibinfo {pages} {4807} (\bibinfo {year} {2022})}\BibitemShut
  {NoStop}%
\bibitem [{\citenamefont {Denisov}\ \emph {et~al.}(2023)\citenamefont
  {Denisov}, \citenamefont {Fuchs}, \citenamefont {Oh},\ and\ \citenamefont
  {Petta}}]{2302.07949}%
  \BibitemOpen
  \bibfield  {author} {\bibinfo {author} {\bibfnamefont {A.~O.}\ \bibnamefont
  {Denisov}}, \bibinfo {author} {\bibfnamefont {G.}~\bibnamefont {Fuchs}},
  \bibinfo {author} {\bibfnamefont {S.~W.}\ \bibnamefont {Oh}},\ and\ \bibinfo
  {author} {\bibfnamefont {J.~R.}\ \bibnamefont {Petta}},\ }\href@noop {} {\
  (\bibinfo {year} {2023})},\ \Eprint {https://arxiv.org/abs/arXiv:2302.07949}
  {arXiv:2302.07949} \BibitemShut {NoStop}%
\bibitem [{\citenamefont {Zajac}\ \emph {et~al.}(2016)\citenamefont {Zajac},
  \citenamefont {Hazard}, \citenamefont {Mi}, \citenamefont {Nielsen},\ and\
  \citenamefont {Petta}}]{PhysRevApplied.6.054013}%
  \BibitemOpen
  \bibfield  {author} {\bibinfo {author} {\bibfnamefont {D.~M.}\ \bibnamefont
  {Zajac}}, \bibinfo {author} {\bibfnamefont {T.~M.}\ \bibnamefont {Hazard}},
  \bibinfo {author} {\bibfnamefont {X.}~\bibnamefont {Mi}}, \bibinfo {author}
  {\bibfnamefont {E.}~\bibnamefont {Nielsen}},\ and\ \bibinfo {author}
  {\bibfnamefont {J.~R.}\ \bibnamefont {Petta}},\ }\href
  {https://doi.org/10.1103/PhysRevApplied.6.054013} {\bibfield  {journal}
  {\bibinfo  {journal} {Phys. Rev. Appl.}\ }\textbf {\bibinfo {volume} {6}},\
  \bibinfo {pages} {054013} (\bibinfo {year} {2016})}\BibitemShut {NoStop}%
\bibitem [{\citenamefont {Weinreb}\ \emph {et~al.}(2009)\citenamefont
  {Weinreb}, \citenamefont {Bardin}, \citenamefont {Mani},\ and\ \citenamefont
  {Jones}}]{doi:10.1063/1.3103939}%
  \BibitemOpen
  \bibfield  {author} {\bibinfo {author} {\bibfnamefont {S.}~\bibnamefont
  {Weinreb}}, \bibinfo {author} {\bibfnamefont {J.}~\bibnamefont {Bardin}},
  \bibinfo {author} {\bibfnamefont {H.}~\bibnamefont {Mani}},\ and\ \bibinfo
  {author} {\bibfnamefont {G.}~\bibnamefont {Jones}},\ }\href
  {https://doi.org/10.1063/1.3103939} {\bibfield  {journal} {\bibinfo
  {journal} {Rev. Sci. Instrum.}\ }\textbf {\bibinfo {volume} {80}},\ \bibinfo
  {pages} {044702} (\bibinfo {year} {2009})}\BibitemShut {NoStop}%
\bibitem [{\citenamefont {Jung}\ \emph {et~al.}(2012)\citenamefont {Jung},
  \citenamefont {Schroer}, \citenamefont {Petersson},\ and\ \citenamefont
  {Petta}}]{doi:10.1063/1.4729469}%
  \BibitemOpen
  \bibfield  {author} {\bibinfo {author} {\bibfnamefont {M.}~\bibnamefont
  {Jung}}, \bibinfo {author} {\bibfnamefont {M.~D.}\ \bibnamefont {Schroer}},
  \bibinfo {author} {\bibfnamefont {K.~D.}\ \bibnamefont {Petersson}},\ and\
  \bibinfo {author} {\bibfnamefont {J.~R.}\ \bibnamefont {Petta}},\ }\href
  {https://doi.org/10.1063/1.4729469} {\bibfield  {journal} {\bibinfo
  {journal} {Appl. Phys. Lett.}\ }\textbf {\bibinfo {volume} {100}},\ \bibinfo
  {pages} {253508} (\bibinfo {year} {2012})}\BibitemShut {NoStop}%
\bibitem [{\citenamefont {Ansaloni}\ \emph {et~al.}(2020)\citenamefont
  {Ansaloni}, \citenamefont {Chatterjee}, \citenamefont {Bohuslavskyi},
  \citenamefont {Bertrand}, \citenamefont {Hutin}, \citenamefont {Vinet},\ and\
  \citenamefont {Kuemmeth}}]{Ansaloni2020}%
  \BibitemOpen
  \bibfield  {author} {\bibinfo {author} {\bibfnamefont {F.}~\bibnamefont
  {Ansaloni}}, \bibinfo {author} {\bibfnamefont {A.}~\bibnamefont
  {Chatterjee}}, \bibinfo {author} {\bibfnamefont {H.}~\bibnamefont
  {Bohuslavskyi}}, \bibinfo {author} {\bibfnamefont {B.}~\bibnamefont
  {Bertrand}}, \bibinfo {author} {\bibfnamefont {L.}~\bibnamefont {Hutin}},
  \bibinfo {author} {\bibfnamefont {M.}~\bibnamefont {Vinet}},\ and\ \bibinfo
  {author} {\bibfnamefont {F.}~\bibnamefont {Kuemmeth}},\ }\href
  {https://doi.org/10.1038/s41467-020-20280-3} {\bibfield  {journal} {\bibinfo
  {journal} {Nat. Commun.}\ }\textbf {\bibinfo {volume} {11}},\ \bibinfo
  {pages} {6399} (\bibinfo {year} {2020})}\BibitemShut {NoStop}%
\bibitem [{\citenamefont {Croot}\ \emph {et~al.}(2019)\citenamefont {Croot},
  \citenamefont {Pauka}, \citenamefont {Jarratt}, \citenamefont {Lu},
  \citenamefont {Gossard}, \citenamefont {Watson}, \citenamefont {Gardner},
  \citenamefont {Fallahi}, \citenamefont {Manfra},\ and\ \citenamefont
  {Reilly}}]{PhysRevApplied.11.064027}%
  \BibitemOpen
  \bibfield  {author} {\bibinfo {author} {\bibfnamefont {X.}~\bibnamefont
  {Croot}}, \bibinfo {author} {\bibfnamefont {S.}~\bibnamefont {Pauka}},
  \bibinfo {author} {\bibfnamefont {M.}~\bibnamefont {Jarratt}}, \bibinfo
  {author} {\bibfnamefont {H.}~\bibnamefont {Lu}}, \bibinfo {author}
  {\bibfnamefont {A.}~\bibnamefont {Gossard}}, \bibinfo {author} {\bibfnamefont
  {J.}~\bibnamefont {Watson}}, \bibinfo {author} {\bibfnamefont
  {G.}~\bibnamefont {Gardner}}, \bibinfo {author} {\bibfnamefont
  {S.}~\bibnamefont {Fallahi}}, \bibinfo {author} {\bibfnamefont
  {M.}~\bibnamefont {Manfra}},\ and\ \bibinfo {author} {\bibfnamefont
  {D.}~\bibnamefont {Reilly}},\ }\href
  {https://doi.org/10.1103/PhysRevApplied.11.064027} {\bibfield  {journal}
  {\bibinfo  {journal} {Phys. Rev. Appl.}\ }\textbf {\bibinfo {volume} {11}},\
  \bibinfo {pages} {064027} (\bibinfo {year} {2019})}\BibitemShut {NoStop}%
\bibitem [{\citenamefont {Hornibrook}\ \emph {et~al.}(2014)\citenamefont
  {Hornibrook}, \citenamefont {Colless}, \citenamefont {Mahoney}, \citenamefont
  {Croot}, \citenamefont {Blanvillain}, \citenamefont {Lu}, \citenamefont
  {Gossard},\ and\ \citenamefont {Reilly}}]{doi:10.1063/1.4868107}%
  \BibitemOpen
  \bibfield  {author} {\bibinfo {author} {\bibfnamefont {J.~M.}\ \bibnamefont
  {Hornibrook}}, \bibinfo {author} {\bibfnamefont {J.~I.}\ \bibnamefont
  {Colless}}, \bibinfo {author} {\bibfnamefont {A.~C.}\ \bibnamefont
  {Mahoney}}, \bibinfo {author} {\bibfnamefont {X.~G.}\ \bibnamefont {Croot}},
  \bibinfo {author} {\bibfnamefont {S.}~\bibnamefont {Blanvillain}}, \bibinfo
  {author} {\bibfnamefont {H.}~\bibnamefont {Lu}}, \bibinfo {author}
  {\bibfnamefont {A.~C.}\ \bibnamefont {Gossard}},\ and\ \bibinfo {author}
  {\bibfnamefont {D.~J.}\ \bibnamefont {Reilly}},\ }\href
  {https://doi.org/10.1063/1.4868107} {\bibfield  {journal} {\bibinfo
  {journal} {Appl. Phys. Lett.}\ }\textbf {\bibinfo {volume} {104}},\ \bibinfo
  {pages} {103108} (\bibinfo {year} {2014})}\BibitemShut {NoStop}%
\end{thebibliography}%

\end{document}